\documentclass{article}
\def\Ref#1{(\ref{#1})}
\usepackage{amsmath,MnSymbol}
\usepackage{cite}
\newcommand{\st}{\mathrm{st}}
\newcommand{\rd}{\mathrm{d}}
\newcommand{\ri}{\mathrm{i}}
\begin{document}
\begin{titlepage}
\begin{center}
{\large \textbf{Phase transitions in a reaction-diffusion model on a line with boundaries}}

\vspace{2\baselineskip}
{\sffamily Mohammad~Khorrami~\footnote{e-mail: mamwad@mailaps.org},
Amir~Aghamohammadi~\footnote{e-mail: mohamadi@alzahra.ac.ir}.}

\vspace{2\baselineskip}
{\it Department of Physics, Alzahra University, Tehran 19938-93973, Iran}
\end{center}
\vspace{4\baselineskip}

\begin{abstract}
\noindent A one-dimensional model on a line of the length $L$ is investigated,
which involves particle diffusion as well as single particle annihilation.
There are also creation and annihilation at the boundaries. The static and
dynamical behaviors of the system are studied. It is seen that the system
could exhibit a dynamical phase transition. For small drift velocities, the relaxation
time does not depend on the absorbtion rates at the boundaries. This is the
fast phase. For large velocities, the smaller of the absorbtion rates at boundaries
enter the relaxation rate and makes it longer. This is the slow phase.
Finally, the effect of a random particle creation in the bulk is also investigated.
\end{abstract}
\end{titlepage}
\newpage
\section{Introduction}
Most analytical studies on reaction diffusion models are focused on
low-dimensional systems. Among these studies, one dimensional models
have the main contribution \cite{ScR,HHL1,MHP2,VPriv}. This is mainly
due to the fact that one dimensional models are easier to investigate,
but at the same time studying such models sheds light on systems far
from equilibrium.

Asymmetric random walk in the continuum limit leads to a diffusion equation
together with a drift (\cite{bAH}, for example), where the time derivative
of the density contains the second derivative of the density with respect
to the position (the diffusion term) and a term proportional to the first
derivative of the density with respect to the position (the drift term). If
in addition to diffusion there are also particle generation, and particle
annihilation, the evolution equation for the density would be the Fisher
equation or the Kolmogorov-Petrovsky-Piskounov equation, where the
time derivative of the density contains a creation term proportional to
the density as well as an annihilation term proportional to the density
squared \cite{Fis, KPP}. In the case of diffusion-limited coalescence,
$A+A\rightleftharpoons A$, one may also have Fisher waves solutions \cite{bA2}.

A simple one-dimensional reaction-diffusion model is a model of some
particles diffusing with a bias. In \cite{RAK}, a continuum version of
such a model (a voting model) in a $D$-dimensional region was studied,
which contained injection and extraction at the boundary as well. Such
a model is in fact a diffusion equation combined with a drift velocity.
The stationary behavior of the system, and the dominant way of relaxation of
the system toward its stationary state was studied there. It was shown that
the system exhibits a static phase transition as well as a dynamical
phase transition. It was shown that the static phase transition is induced
by the drift velocity only, while the dynamical phase transition is a result of
both the drift velocity and boundary conditions.

Adding a random particle creation term to a diffusion, one is arrived at
the Edwards-Wilkinson model. The effect of a drift to the scaling
properties of such a model is discussed in \cite{pr1}. The model
considered there, is on a one dimensional loop. Hence there are
no boundary effects. In \cite{pr2}, the Edwards-Wilkinson model
with Neumann boundary conditions has been investigated, again
focusing on the scaling properties of the system.

Here a one-dimensional reaction-diffusion model on a line of
the length $L$ with open boundaries is addressed. In addition
to the reactions in the bulk, there are also creation and
annihilation (or injection and extraction of particles) at
the boundaries. Among other things, the possible existence of phase
transitions is studied. By phase transition, it is meant a discontinuity
in some behavior of the system with respect to its parameters. The static
phase transition is a discontinuity in the stationary (large time)
profile of the system, and the dynamical phase transition corresponds to
a discontinuous change in the behavior of the relaxation time of
the system toward its stationary state. Finally, a random particle creation
is added to the system. The result would be similar to the Edwards-Wilkinson
model plus a drift and an annihilation term, with general (linear)
boundary condition. Specifically, the one-and two point functions
are obtained, explicitly in the thermodynamic limit.

The scheme of the paper is as follows. In section 2, the model is
introduced.  In section 3, the time-independent solution is
studied. It is shown that in the thermodynamic limit the boundaries
essentially act independent of each other, and the stationary solution
is non-vanishing only near boundaries. The conditions for the existence
of a physical stationary solution are also obtained. Section 4 is devoted
to investigating the dynamical behavior of the system. It is seen that
the system has two phases, the fast phase and the slow phases, and it
can exhibit a dynamical phase transition. For large absorbtion rates,
the relaxation time does not depend on the absorbtion
rates at the boundaries. This is the fast phase. For small absorbtion rates,
the smaller of the absorbtion rates at boundaries enter the relaxation rate
and makes it longer. This is the slow phase. In section 5, a random particle
creation is added to the model, and its effect on the particle density and
two-point correlations is investigated. Section 6 is devoted to the concluding
remarks.
\section{The model}
Consider a one-dimensional line of the length $L$, on which particles
of a single type diffuse, drift, and undergo single particle annihilation.
At the boundaries ($x=0$ and $x=L$), there is creation and annihilation
as well. The density of particles, denoted by $\rho$, satisfies
the differential equation
\begin{equation}\label{Hm01.1}
D_0\rho=D^2\rho-u\,D\rho-a\,\rho,
\end{equation}
where $D_0$ and $D$ denote differentiation with respect to $t$ (time)
and $x$ (position), respectively, the diffusion constant has been
absorbed in a redefinition of time, $u$ is the drift velocity, and
$a$ is the annihilation parameter. The evolution equation (\ref{Hm01.1})
is the most general linear local (in position) translation invariant
equation which ensures that the density remains nonnegative if one
begins with a nonnegative initial condition \cite{HmIJTP}.
The boundary conditions are
\begin{alignat}{2}\label{Hm01.2}
D\rho&=-\alpha+\beta\,\rho,&\qquad x&=0,\nonumber\\
D\rho&=\alpha'-\beta'\,\rho,&\qquad x&=L,
\end{alignat}
where $\alpha$ and $\beta$ are respectively the creation and annihilation
parameters at $x=0$, while $\alpha'$ and $\beta'$ are similar parameters
at $x=L$. $a$, $\alpha$, $\beta$, $\alpha'$, and $\beta'$ are assumed to be nonnegative.
It is easy to see that the total number of particles evolve in time like
\begin{equation}\label{Hm01.3}
D_0\left(\int_0^L\rd x\;\rho\right)=\alpha'-\beta'\,\rho(L)+\alpha-\beta\,\rho(0)
-u\,\rho(L)+u\,\rho(0)-a\,\int_0^L\rd x\;\rho.
\end{equation}
As a result of the existence of boundaries, the overall evolution is not translation invariant,
although the differential equation is.

\section{The stationary behavior}
The stationary solution to \Ref{Hm01.1} and \Ref{Hm01.2} is denoted by
$\rho^\st$. It is seen that
\begin{equation}\label{Hm01.4}
\rho^\st(x)=A\,\exp\left[\left(\frac{u}{2}-\gamma\right)\,x\right]+
A'\,\exp\left[\left(\frac{u}{2}+\gamma\right)\,(x-L)\right],
\end{equation}
where
\begin{equation}\label{Hm01.5}
\gamma:=\sqrt{a+\frac{u^2}{4}},
\end{equation}
and the constants $A$ and $A'$ satisfy
\begin{align}\label{Hm01.6}
(\tilde\beta+\gamma)\,A+\left\{(\tilde\beta-\gamma)\,
\exp\left[-\left(\frac{u}{2}+\gamma\right)\,L\right]\right\}\,A'&=\alpha,\nonumber\\
\left\{(\tilde\beta'-\gamma)\,\exp\left[\left(\frac{u}{2}-\gamma\right)\,L\right]\right\}
\,A+(\tilde\beta'+\gamma)\,A'&=\alpha',
\end{align}
where
\begin{align}\label{Hm01.7}
\tilde\beta&:=\beta-\frac{u}{2},\nonumber\\
\tilde\beta'&:=\beta'+\frac{u}{2}.
\end{align}
A negative annihilation parameter ($a$) is equivalent to particle creation.
A negative $a$ can make the stationary solution obtained above unacceptable.
The point is that $\rho^\st$ should be nonnegative. In the thermodynamic limit
($L\to\infty$), in order that this be the case $\gamma$ should be real.
If $\gamma$ is imaginary then $\rho^\st$ is an exponential times
a harmonic oscillation, which changes sign. So one criterion that there is
a stationary solution which is the limit of the density at large times, in
the thermodynamic limit is that $\gamma$ be real. This can be written as
\begin{equation}\label{Hm01.8}
a> -\frac{u^2}{4}.
\end{equation}
Assuming that this is the case, and denoting by $\gamma$ the positive root,
three cases occur (regarding the thermodynamic limit).\\
\textbf{case i:}
\begin{equation}\label{Hm01.9}
\gamma>\left|\frac{u}{2}\right|.
\end{equation}
Then, in the thermodynamic limit,
\begin{align}\label{Hm01.10}
A&=\frac{\alpha}{\tilde\beta+\gamma},\nonumber\\
A'&=\frac{\alpha'}{\tilde\beta'+\gamma}.
\end{align}
This means that in the thermodynamic limit the boundaries essentially
act independent of each other, and the stationary solution is
non-vanishing only near boundaries:
\begin{alignat}{2}\label{Hm01.11}
\rho^\st(x)&\approx \frac{\alpha}{\tilde\beta+\gamma}\,
\exp\left[-\left(\gamma-\frac{u}{2}\right)\,x\right],&\qquad &x\ll L,\nonumber\\
\rho^\st(x)&\approx \frac{\alpha'}{\tilde\beta'+\gamma}\,
\exp\left[-\left(\gamma+\frac{u}{2}\right)\,(L-x)\right],&\qquad &(L-x)\ll L,\nonumber\\
\rho^\st(x)&\approx 0,&\qquad &\mathrm{otherwise}.
\end{alignat}
Such solutions are acceptable, only if both $(\tilde\beta+\gamma)$ and
$(\tilde\beta'+\gamma)$ are positive, to ensure the positivity of $\rho^\st$.\\[\baselineskip]
\textbf{case ii:}
\begin{equation}\label{Hm01.12}
\gamma<\frac{u}{2}.
\end{equation}
Then, in the thermodynamic limit,
\begin{align}\label{Hm01.13}
A&=\frac{\alpha}{\tilde\beta+\gamma},\nonumber\\
A'&=-\frac{\alpha}{\tilde\beta+\gamma}\,\exp\left[\left(\frac{u}{2}-\gamma\right)\,L\right].
\end{align}
So
\begin{equation}\label{Hm01.14}
\rho^\st(x)\approx \frac{\alpha}{\tilde\beta+\gamma}\,
\exp\left[-\left(\gamma-\frac{u}{2}\right)\,x\right],\qquad (L-x)\nll L.
\end{equation}
The static solution rises exponentially, from $x=0$ up to points near
the boundary $x=L$. For the static density to be positive,
$(\tilde\beta+\gamma)$ should be positive. In this case,
$(\tilde\beta'+\gamma)$ is already positive, as $u$ is positive.\\[\baselineskip]
\textbf{case iii:}
\begin{equation}\label{Hm01.15}
\gamma<-\frac{u}{2}.
\end{equation}
Then, in the thermodynamic limit,
\begin{align}\label{Hm01.16}
A&=-\frac{\alpha'}{\tilde\beta'+\gamma}\,\exp\left[-\left(\frac{u}{2}+\gamma\right)\,L\right],\nonumber\\
A'&=\frac{\alpha'}{\tilde\beta'+\gamma}.
\end{align}
So
\begin{equation}\label{Hm01.17}
\rho^\st(x)\approx \frac{\alpha'}{\tilde\beta'+\gamma}\,
\exp\left[-\left(\gamma+\frac{u}{2}\right)\,(L-x)\right],\qquad x\nll L.
\end{equation}
The static solution rises exponentially, from $x=L$ up to points near
the boundary $x=0$. For the static density to be positive,
$(\tilde\beta'+\gamma)$ should be positive. In this case,
$(\tilde\beta+\gamma)$ is already positive, as $u$ is negative.

To summarize, in order to have a positive stationary density
(\ref{Hm01.8}) should hold, as well as
\begin{align}\label{Hm01.18}
\tilde\beta+\gamma > 0,\nonumber\\
\tilde\beta'+\gamma>0.
\end{align}
(\ref{Hm01.8}) and (\ref{Hm01.18}) are the conditions that (in the thermodynamic
limit) there is a stationary solution which is the limit of the density at large times.
They can be written as
\begin{equation}\label{Hm01.19}
a>\max\left(-\frac{u^2}{4}+\tilde\beta^2,-\frac{u^2}{4}+\tilde\beta'^2\right).
\end{equation}
If this is satisfied, even with particle creation the drift velocity and
the annihilation rates at the boundaries are high enough to ensure the rapid
annihilation of particles at the boundaries, so that
the particle density does not grow exponentially at large times. There is
a difference between the cases of particle creation and particle annihilation.
If there is particle creation, then the stationary density grows exponentially
in $x$, towards the direction of $u$, so that it becomes
very large near one of the boundaries, although the density does not grow exponentially
in time, meaning that for each $x$ it tends to the finite value $\rho^\st(x)$ as
time tends to infinity. If there is particle annihilation, then $\rho^\st$
is bounded (for all values of $x$), and essentially is nonzero only near the boundaries.
\section{The dynamic behavior}
The dynamic solution $(\rho-\rho^\st)$ satisfies
\begin{equation}\label{Hm01.20}
D_0(\rho-\rho^\st)=H\,(\rho-\rho^\st),
\end{equation}
where $H$ is defined through
\begin{equation}\label{Hm01.21}
H\,\psi=D^2\,\psi-u\,D\,\psi-a\,\psi,
\end{equation}
and the boundary conditions
\begin{alignat}{2}\label{Hm01.22}
D\,\psi&=\beta\,\psi,&\qquad x&=0,\nonumber\\
D\,\psi&=-\beta'\,\psi,&\qquad x&=L.
\end{alignat}
Putting the eigensolution ansatz $\psi_E(x)\,\exp(E\,t)$ in (\ref{Hm01.20}),
one arrives at
\begin{equation}\label{Hm01.23}
D^2\psi_E-u\,D\psi_E-(a+E)\,\psi_E=0,
\end{equation}
Of course, $\psi_E$ satisfies the boundary conditions (\ref{Hm01.22}).
The solution to these is
\begin{equation}\label{Hm01.24}
\psi_E(x)=\exp\left(\frac{u}{2}\,x\right)\,\{B\,\exp(-q\,x)+B'\,\exp[q\,(x-L)]\},
\end{equation}
where $q$ is a constant with a nonnegative real part, satisfying
\begin{equation}\label{Hm01.25}
q^2=\gamma^2+E,
\end{equation}
and $B$ and $B'$ are other constants satisfying
\begin{align}\label{Hm01.26}
(-q-\tilde\beta)\,B+[(q-\tilde\beta)\,\exp(-q\,L)]\,B'&=0,\nonumber\\
[(-q+\tilde\beta')\,\exp(-q\,L)]\,B+(q+\tilde\beta')\,B'&=0.
\end{align}
It is seen that in the thermodynamic limit, if the real part of $q$ is non-vanishing then
\begin{equation}\label{Hm01.27}
(-q-\tilde\beta)\,(q+\tilde\beta')=0,
\end{equation}
which results in a real solution for $q$ provided that solution is positive. Denoting this solution for
$q$ by $q_\mathrm{r}$, it is seen that
\begin{equation}\label{Hm01.28}
q_\mathrm{r}=\begin{cases}
-\tilde\beta',&u<(-2\,\beta')\\
-\tilde\beta,&u>(2\,\beta)\end{cases}
\end{equation}
and that there is no real solution for $q$ if $(-2\,\beta')<u<(2\,\beta)$.
Other solutions for $q$ are pure imaginary. So all values of $E$ are real. Denoting the
largest value of $E$ by $E_\mathrm{m}$, it is seen that
\begin{equation}\label{Hm01.29}
E_\mathrm{m}=\begin{cases}
-\gamma^2+\tilde\beta'^2,&u<(-2\,\beta')\\
-\gamma^2,&(-2\,\beta')<u<(2\,\beta)\\
-\gamma^2+\tilde\beta^2,&(2\,\beta)<u
\end{cases}
\end{equation}
The relaxation time of the system is determined by $E_\mathrm{m}$:
\begin{equation}\label{Hm01.30}
\tau=-\frac{1}{E_\mathrm{m}}.
\end{equation}
So there is a dynamical phase transition. For small drift velocities,
the relaxation time does not depend on the absorbtion rates at the boundaries.
This is the fast phase. For large velocities, the smaller of the absorbtion rates
at boundaries enter the relaxation rate and makes it longer. This is the slow phase.

Let us again address the case of particle creation (negative $a$). It is seen
that $E_\mathrm{m}$ is negative provided (\ref{Hm01.19}) is satisfied. So
the system has a stationary state relaxing exponentially towards it, provided
(\ref{Hm01.19}) is satisfied.
\section{The stochastic evolution}
The evolution equation (\ref{Hm01.1}) for the density $\rho$
is a deterministic one. One way to introduce stochasticity in
the evolution for $\rho$, is to introduce some random particle
creation. This could be done by adding a noise $\eta$ to
the right-hand side of (\ref{Hm01.1}), so that the evolution
for the stochastic density $\hat\rho$ is
\begin{equation}\label{Hm01.31}
D_0\hat\rho=D^2\hat\rho-u\,D\hat\rho-a\,\hat\rho+\eta,
\end{equation}
where $\eta$ is a stochastic function. In order that no initial
condition for $\hat\rho$ results in a solution for $\hat\rho$
that sometime and somewhere is negative, the probability
that $\eta$ be negative should be zero. However, one can still
take the values of $\eta$ at different times or positions
independent. Assuming so, one has
\begin{align}\label{Hm01.32}
\langle\eta(t,x)\rangle&=f_1(t,x),\\ \label{Hm01.33}
\langle\eta(t,x)\,\eta(t',x')\rangle
-\langle\eta(t,x)\rangle\,\langle\eta(t',x')\rangle&=f_2(t,x)\,\delta(t-t')\,\delta(x-x'),
\end{align}
where $f_1$ and $f_2$ are positive. The solution to (\ref{Hm01.31}) satisfies
\begin{equation}\label{Hm01.34}
\hat\rho=\rho+\rho^\mathrm{n},
\end{equation}
where $\rho$ is the solution to (\ref{Hm01.1}) and the (inhomogeneous)
boundary conditions and the initial conditions, while $\rho^\mathrm{n}$
(which is the stochastic part of $\hat\rho$) satisfies
\begin{equation}\label{Hm01.35}
D_0\rho^\mathrm{n}=H\,\rho^\mathrm{n}+\eta,
\end{equation}
subject to the boundary conditions
\begin{alignat}{2}\label{Hm01.36}
D\rho^\mathrm{n}&=\beta\,\rho^\mathrm{n},&\qquad x&=0,\nonumber\\
D\rho^\mathrm{n}&=-\beta'\,\rho^\mathrm{n},&\qquad x&=L,
\end{alignat}
and initial value being zero.
The solution to
(\ref{Hm01.35}) and (\ref{Hm01.36}) with zero initial value is
\begin{equation}\label{Hm01.37}
\rho^\mathrm{n}(t,x)=\int_{-\infty}^\infty\rd t'\int_0^L\rd x'\;G(t-t';x,x')\,\eta(t',x'),
\end{equation}
where the Green's function $G$ satisfies
\begin{equation}\label{Hm01.38}
(D_0 G)(t;x,x')=(H\,G)(t;x,x')+\delta(t)\,\delta(x-x'),
\end{equation}
subject to the boundary conditions
\begin{alignat}{2}\label{Hm01.39}
D G&=\beta\,G,&\qquad x&=0,\nonumber\\
D G&=-\beta'\,G,&\qquad x&=L,
\end{alignat}
and zero initial condition. To prove (\ref{Hm01.37}) with the above
Green's function is in fact the solution with proper boundary conditions
and initial conditions, one notices that acting with
$(D-\beta)$ on $\rho^\mathrm{n}$ results in acting with
$(D-\beta)$ on $G$ in the right hand side of (\ref{Hm01.37}).
The same is true for $(D+\beta')$. So if the boundary conditions
(\ref{Hm01.39}) hold, so do the boundary conditions
(\ref{Hm01.36}). Regarding the initial condition, one notices
that the zero initial condition for the Green's function,
together with the evolution equation (\ref{Hm01.38}),
results in the fact that $G(t;x,x')$ is zero for negative
$t$. So the integration region on $t'$ in the right hand side of
(\ref{Hm01.37}) is essentially $(-\infty,t]$. Hence for $t$
to $-\infty$, the integration region is zero and $\rho^\mathrm{n}$
vanishes, meaning that it does satisfy the (zero) initial condition.
Finally, acting with $(D_0-H)$ on $\rho^\mathrm{n}$, results in
acting by $(D_0-H)$ on the right-hand side of (\ref{Hm01.37}).
The resulting integrand is the product of delta functions in the
right-hand side of (\ref{Hm01.38}), using which the integrations
are done and (\ref{Hm01.35}) is recovered.

Defining
\begin{align}\label{Hm01.40}
C_1(t,x)&:=\langle\hat\rho(t,x)\rangle,\\ \label{Hm01.41}
C_2(t_1,t_2;x_1,x_2)&:=\langle\hat\rho(t_1,x_1)\,\hat\rho(t_2,x_2)\rangle-
\langle\hat\rho(t_1,x_1)\rangle\,\langle\hat\rho(t_2,x_2)\rangle,
\end{align}
using (\ref{Hm01.32}), (\ref{Hm01.33}), and (\ref{Hm01.39}),
one arrives at
\begin{align}\label{Hm01.42}
C_1(t,x)&=\rho(t,x)+\int_{-\infty}^\infty\rd t'\int_0^L\rd x'\;G(t-t';x,x')\,f_1(t',x'),\\ \label{Hm01.43}
C_2(t_1,t_2;x_1,x_2)&=\int_{-\infty}^\infty\rd t'\int_0^L\rd x'\;
G(t_1-t';x_1,x')\,G(t_2-t';x_2,x')\,f_2(t',x').
\end{align}

One has
\begin{equation}\label{Hm01.44}
G(t;x,x')=\sum_E\,\psi^*_E(x')\,\psi_E(x)\,\exp(E\,t)\,\Theta(t),
\end{equation}
where
$\psi^*_E$ is the eigenvector of $H^*$ (the pull back of $H$)
with the eigenvalue $E$, normalized so that
\begin{equation}\label{Hm01.45}
\int_0^L\rd x\;\psi^*_E(x)\,\psi_E(x)=1.
\end{equation}
$H^*$ satisfies
\begin{equation}\label{Hm01.46}
\int_0^L\rd x\;(H^*\,\psi^*)(x)\,\psi(x)=\int_0^L\rd x\;\psi^*(x)\,(H\,\psi)(x).
\end{equation}
So,
\begin{equation}\label{Hm01.47}
H^*\,\psi^*=D^2\psi^*+u\,D\psi^*-a\,\psi^*,
\end{equation}
where $\psi^*$ should satisfy
\begin{equation}\label{Hm01.48}
\left[\psi^*\,\frac{\rd\psi}{\rd x}-\frac{\rd\psi^*}{\rd x}\,\psi-u\,\psi^*\,\psi\right]^L_0=0.
\end{equation}
Using the boundary conditions (\ref{Hm01.22}) for $\psi$, it is seen
that the proper boundary conditions for $\psi^*$ are
\begin{alignat}{2}\label{Hm01.49}
D\psi^*&=(\beta-u)\,\psi^*,&\qquad x&=0,\nonumber\\
D\psi^*&=-(\beta'+u)\,\psi^*,&\qquad x&=L.
\end{alignat}
Comparing (\ref{Hm01.21}) and (\ref{Hm01.22}) with (\ref{Hm01.47}) and
(\ref{Hm01.49}), respectively, it is seen that $H^*$ is obtained from
$H$ by a change of $(u,\beta,\beta')\to(u^*,\beta^*,\beta'^*)$,
where
\begin{align}\label{Hm01.50}
u^*&=-u,\nonumber\\
\beta^*-\frac{u^*}{2}&=\beta-\frac{u}{2},\nonumber\\
\beta'^*+\frac{u^*}{2}&=\beta'+\frac{u}{2}.
\end{align}
One thus arrives at
\begin{equation}\label{Hm01.51}
\psi^*_E(x)=\exp\left(-\frac{u}{2}\,x\right)\,\{B\,\exp(-q\,x)+B'\,\exp[q\,(x-L)]\},
\end{equation}
with the same values of $q$, $B$, and $B'$ appeared in (\ref{Hm01.24}).
The normalization condition (\ref{Hm01.45}) then reads
\begin{equation}\label{Hm01.52}
(B^2+B'^2)\,\frac{\sinh q\,L}{q\,L}+2\,B\,B'=\frac{\exp(q\,L)}{L}.
\end{equation}
\subsection{The thermodynamic limit}
In the thermodynamic limit ($L\to\infty$), the difference between
two consecutive values of imaginary $q$'s become behaves like
$(\ri\,\pi/L)$. Keeping $x$ finite, using (\ref{Hm01.52}), in which the first term on
the right hand side becomes negligible, and the first
boundary condition in (\ref{Hm01.26}), one arrives at
\begin{align}\label{Hm01.53}
B&=\frac{\ri}{\sqrt{2\,L}}\,\exp\left(-\ri\,\tan^{-1}\frac{k}{\tilde\beta}\right),\nonumber\\
B'\,\exp(-q\,L)&=\frac{-\ri}{\sqrt{2\,L}}\,\exp\left(\ri\,\tan^{-1}\frac{k}{\tilde\beta}\right),
\end{align}
where
\begin{equation}\label{Hm01.54}
q=:\ri\,k.
\end{equation}
So,
\begin{equation}\label{Hm01.55}
\psi_E(x)=\sqrt{\frac{2}{L}}\,\exp\left(\frac{u}{2}\,x\right)\,
\sin\left(k\,x+\tan^{-1}\frac{k}{\tilde\beta}\right),
\qquad E<-\gamma^2,
\end{equation}
where
\begin{equation}\label{Hm01.56}
E=-\gamma^2-k^2.
\end{equation}
The Green's function for finite $x$ would have a contribution from
a real positive value of $q$, if $\tilde\beta$ is negative. One has
\begin{equation}\label{Hm01.57}
\psi_E(x)=\sqrt{-2\,\tilde\beta}\,\exp(\beta\,x),\qquad E=-\gamma^2+\tilde\beta^2.
\end{equation}
So the Green's function at the thermodynamic limit, for finite
values of $x$ and $x'$, satisfies
\begin{align}\label{Hm01.58}
G(t;x,x')&=\exp\left[\frac{u}{2}\,(x-x')\right]\,\Bigg\{
-2\,\tilde\beta\,\exp[\tilde\beta\,(x+x')]
\,\exp[-(\gamma^2-\tilde\beta^2)\,t]\,
\Theta(-\tilde\beta)\nonumber\\
&\qquad+\frac{2}{\pi}\,\int_0^\infty\rd k\;
\sin\left(k\,x+\tan^{-1}\frac{k}{\tilde\beta}\right)\,
\sin\left(k\,x'+\tan^{-1}\frac{k}{\tilde\beta}\right)\nonumber\\
&\qquad\quad\times\exp[-(\gamma^2+k^2)\,t]\Bigg\}\,
\Theta(t).
\end{align}
Defining
\begin{align}\label{Hm01.59}
\xi&:=x-x',\nonumber\\
X&:=x+x',
\end{align}
one arrives at
\begin{equation}\label{Hm01.60}
G(t;x,x')=\exp\left(\frac{u}{2}\,\xi-\gamma^2\,t\right)\,\Theta(t)\,
\tilde G(t;x,x'),
\end{equation}
where
\begin{align}\label{Hm01.61}
\tilde G(t;x,x')&=
\tilde G_\mathrm{b}(t,X)+\tilde G_\mathrm{f}(t,\xi)+\tilde G_\mathrm{i}(t,X),\nonumber\\
\tilde G_\mathrm{b}(t,X)&=-2\,\tilde\beta\,\exp(\tilde\beta\,X)
\,\exp(\tilde\beta^2\,t)\,\Theta(-\tilde\beta),\nonumber\\
\tilde G_\mathrm{f}(t,\xi)&=\frac{1}{\pi}\,\int_0^\infty\rd k\;\cos(k\,\xi)\,
\exp(-k^2\,t),\nonumber\\
\tilde G_\mathrm{i}(t,X)&=-\frac{1}{\pi}\,\int_0^\infty\rd k\;
\cos\left(k\,X+2\,\tan^{-1}\frac{k}{\tilde\beta}\right)\,
\exp(-k^2\,t).
\end{align}
It is easily seen that
\begin{equation}\label{Hm01.62}
\tilde G_\mathrm{f}(t,\xi)=\frac{1}{\sqrt{4\,\pi\,t}}\,\exp\left(-\frac{\xi^2}{4\,t}\right).
\end{equation}
For $\tilde G_\mathrm{i}$, one has
\begin{align}\label{Hm01.63}
\tilde G_\mathrm{i}(t,X)&=-\frac{1}{2\,\pi}\int_{-\infty}^\infty\rd k\;
\exp\left(\ri\,k\,X+2\,\ri\,\tan^{-1}\frac{k}{\tilde\beta}-k^2\,t\right),\nonumber\\
&=\frac{1}{2\,\pi}\int_{-\infty}^\infty\rd k\;\frac{\ri\,k+\tilde\beta}{\ri\,k-\tilde\beta}\;
\exp(\ri\,k\,X-k^2\,t),\nonumber\\
&=\frac{1}{\sqrt{4\,\pi\,t}}\,\exp\left(-\frac{X^2}{4\,t}\right)+
\frac{\tilde\beta}{\pi}\,\exp(\tilde\beta\,X)\,\tilde G_\mathrm{e}(t,X),
\end{align}
where
\begin{equation}\label{Hm01.64}
\tilde G_\mathrm{e}(t,X):=\int_{-\infty}^\infty\rd k\;
\frac{1}{\ri\,k-\tilde\beta}\;\exp[(\ri\,k-\tilde\beta)\,X-k^2\,t].
\end{equation}
One has,
\begin{align}\label{Hm01.65}
\tilde G_\mathrm{e}(t,X)&=\tilde G_\mathrm{e}(t,0)+\int_0^X\rd w\;\int_{-\infty}^\infty\rd k\;
\exp[(\ri\,k-\tilde\beta)\,w-k^2\,t],\nonumber\\
&=\tilde G_\mathrm{e}(t,0)+\sqrt{\frac{\pi}{t}}\,\exp(\tilde\beta^2\,t)\,
\int_0^X\rd w\;\exp\left[-\frac{1}{4\,t}\,(w+2\,\tilde\beta\,t)^2\right],\nonumber\\
&=\tilde G_\mathrm{e}(t,0)+\pi\,\exp(\tilde\beta^2\,t)\,
\left[\mathrm{erf}\left(\frac{X+2\,\tilde\beta\,t}{2\,\sqrt{t}}\right)
-\mathrm{erf}(\tilde\beta\,\sqrt{t})\right].
\end{align}
For $\tilde G_\mathrm{e}(t,0)$, one notices that if $\tilde\beta$ is
positive, then $\tilde G_\mathrm{e}(t,X)$ tends to zero for $X\to\infty$. So,
\begin{equation}\label{Hm01.66}
\tilde G_\mathrm{e}(t,0)=-\pi\,\exp(\tilde\beta^2\,t)\,[1-\mathrm{erf}(\tilde\beta\,\sqrt{t})],\qquad
\tilde\beta>0.
\end{equation}
As $\tilde G_\mathrm{e}(t,0)$ is odd in $\tilde\beta$, on arrives at
\begin{equation}\label{Hm01.67}
\tilde G_\mathrm{e}(t,0)=-\pi\,\exp(\tilde\beta^2\,t)\,
[\mathrm{sgn}(\tilde\beta)-\mathrm{erf}(\tilde\beta\,\sqrt{t})].
\end{equation}
So,
\begin{align}\label{Hm01.68}
G(t;x,x')&=\exp\left(\frac{u}{2}\,\xi-\gamma^2\,t\right)\,\Theta(t)\,
\Bigg\{\frac{1}{\sqrt{4\,\pi\,t}}\,\left[\exp\left(-\frac{\xi^2}{4\,t}\right)+
\exp\left(-\frac{X^2}{4\,t}\right)\right]\nonumber\\
&\qquad-\tilde\beta\,\exp(\tilde\beta\,X)\,\exp(\tilde\beta^2\,t)\,
\mathrm{erfc}\left(\frac{X+2\,\tilde\beta\,t}{2\,\sqrt{t}}\right)\Bigg\}.
\end{align}
\subsection{Time independent uniform noise}
A special case is when the correlations of noise are
time independent and uniform, so that $f_1$ and $f_2$ in
(\ref{Hm01.32}) and (\ref{Hm01.33}) are positive constants.
In this case, (\ref{Hm01.42}) and (\ref{Hm01.43}) become
\begin{align}\label{Hm01.69}
C_1(t,x)&=\rho(t,x)+f_1\,\int_{-\infty}^\infty\rd t'\int_0^L\rd x'\;G(t-t';x,x'),\\ \label{Hm01.70}
C_2(t_1,t_2;x_1,x_2)&=f_2\int_{-\infty}^\infty\rd t'\int_0^L\rd x'\;
G(t_1-t';x_1,x')\,G(t_2-t';x_2,x').
\end{align}

For the one point function $C_1$, one needs the integral of the Green's function:
\begin{equation}\label{Hm01.71}
G_1(t,x):=\int_{-\infty}^\infty\rd t'\int_0^L\rd x'\;G(t-t';x,x').
\end{equation}
It is seen that $G_1$ is time independent and satisfies
\begin{align}\label{Hm01.72}
(H\,G_1)(x)+1&=0,\\ \label{Hm01.73}
(D\,G_1)(0)&=\beta\,G_1(0),\\ \label{Hm01.74}
(D\,G_1)(L)&=-\beta'\,G_1(L).
\end{align}
The solution to this, is
\begin{equation}\label{Hm01.75}
G_1(x)=\frac{1}{a}+A_1\,\exp\left[\left(\frac{u}{2}-\gamma\right)\,x\right]
+A'_1\,\exp\left[\left(\frac{u}{2}+\gamma\right)\,(x-L)\right],
\end{equation}
with $A_1$ and $A'_1$ satisfying
\begin{align}\label{Hm01.76}
(\tilde\beta+\gamma)\,A+\left\{(\tilde\beta-\gamma)\,
\exp\left[-\left(\frac{u}{2}+\gamma\right)\,L\right]\right\}\,A'&=-\frac{\beta}{a},\nonumber\\
\left\{(\tilde\beta'-\gamma)\,\exp\left[\left(\frac{u}{2}-\gamma\right)\,L\right]\right\}
\,A+(\tilde\beta'+\gamma)\,A'&=-\frac{\beta'}{a},
\end{align}
In the thermodynamic limit, these result in
\begin{align}\label{Hm01.77}
A&=-\frac{\beta}{a\,(\tilde\beta+\gamma)},\nonumber\\
A'&=-\frac{\beta'}{a\,(\tilde\beta'+\gamma)},
\end{align}
so that in the thermodynamic limit and for finite values of x,
\begin{equation}\label{Hm01.78}
G_1(x)=\frac{1}{a}\left\{1-\frac{\beta}{\tilde\beta+\gamma}\,
\exp\left[-\left(\gamma-\frac{u}{2}\right)\,x\right]\right\}.
\end{equation}

For the two point function with noise, one needs $G_2$ with
\begin{equation}\label{Hm01.79}
G_2(t;x_1,x_2):=\int_{-\infty}^\infty\rd s\int_0^L\rd x'\;G(t+s;x_1,x')\,G(s;x_2,x'),
\end{equation}
where (without loss of generality) it is assumed that $t$
(which is in fact $(t_1-t_2)$) is positive. It is seen that $G_2$ satisfies
\begin{align}\label{Hm01.80}
(D_0-H_1)\,G_2&=0,\\ \label{Hm01.81}
(D_0+H_2)\,G_2&=-G,
\end{align}
where $H_i$ is the differential operator $H$ corresponding to the
variable $x_i$. Of course, $G_2$ satisfies the boundary conditions
(\ref{Hm01.22}) for both variables $x_1$ and $x_2$.
Using (\ref{Hm01.80}), one arrives at
\begin{equation}\label{Hm01.82}
G_2(t;x_1,x_2)=\sum_E\,\psi_E(x_1)\,\chi_E(x_2)\,\exp(E\,t),
\end{equation}
where $E$'s are the eigenvalues of $H$. Putting this in
(\ref{Hm01.81}), and using (\ref{Hm01.44}), one arrives at
\begin{equation}\label{Hm01.83}
(H+E)\,\chi_E=-\psi^*_E,
\end{equation}
the solution to which is
\begin{align}\label{Hm01.84}
\chi_E(x)&=-\exp\left(-\frac{u}{2}\,x\right)\,\left\{\frac{B}{u\,(u+2\,q)+2\,E}\,\exp(-q\,x)\right.
\nonumber\\
&\qquad+\left.
\frac{B'}{u\,(u-2\,q)+2\,E}\,\exp[q\,(x-L)]\right\}\nonumber\\
&\quad+\exp\left(\frac{u}{2}\,x\right)\,\{B_2\,\exp(-r\,x)+B'_2\,\exp[r\,(x-L)]\},
\end{align}
where
\begin{equation}\label{Hm01.85}
r=\sqrt{\gamma^2-E},
\end{equation}
and $B_2$ and $B'_2$ should be such that $\chi_E$ satisfy the boundary
conditions \Ref{Hm01.22}, that is
\begin{align}\label{Hm01.86}
(\tilde\beta+r)\,B_2+(\tilde\beta-r)\,B'_2\,\exp(-r\,L)&=
\frac{(\tilde\beta+u+q)\,B}{u\,(u+2\,q)+2\,E}\nonumber\\
&\quad+\frac{(\tilde\beta+u-q)\,B'\,\exp(-q\,L)}{u\,(u-2\,q)+2\,E},\nonumber\\
(-\tilde\beta'+r)\,B_2\,\exp(-r\,L)+(-\tilde\beta'-r)\,B'_2&=
\frac{(-\tilde\beta'+u+q)\,B\,\exp(-q\,L)}{u\,(u+2\,q)+2\,E}\nonumber\\
&\quad+\frac{(-\tilde\beta'+u-q)\,B'}{u\,(u-2\,q)+2\,E},
\end{align}
which in the thermodynamic limit result in
\begin{align}\label{Hm01.87}
B_2&=\frac{1}{\tilde\beta+r}\,\left[
\frac{(\tilde\beta+u+q)\,B}{u\,(u+2\,q)+2\,E}+
\frac{(\tilde\beta+u-q)\,B'\,\exp(-q\,L)}{u\,(u-2\,q)+2\,E}\right],\nonumber\\
B'_2&=-\frac{1}{\tilde\beta'+r}\,\left[
\frac{(-\tilde\beta'+u+q)\,B\,\exp(-q\,L)}{u\,(u+2\,q)+2\,E}+
\frac{(-\tilde\beta'+u-q)\,B'}{u\,(u-2\,q)+2\,E}\right].
\end{align}
So in the thermodynamic limit, and for finite values of $x_1$ and $x_2$, one arrives at
\begin{equation}\label{Hm01.88}
G_2(t;x_1,x_2)=\exp(-\gamma^2\,t)\,[\tilde G_{2\,\mathrm{b}}(t;x_1,x_2)
+\tilde G_{2\,\mathrm{u}}(t;x_1,x_2)],
\end{equation}
where
\begin{align}\label{Hm01.89}
\tilde G_{2\,\mathrm{b}}(t;x_1,x_2)&=\frac{-2\,\tilde\beta}
{u\,(u-2\,\tilde\beta)-2\,\gamma^2+2\,\tilde\beta^2}
\,\exp(\beta\,x_1+\tilde\beta^2\,t)\nonumber\\
&\quad\times\Bigg\{
\frac{u}{\tilde\beta+\sqrt{2\,\gamma^2-\tilde\beta^2}}\,
\exp\left[\left(\frac{u}{2}-\sqrt{2\,\gamma^2-\tilde\beta^2}\right)\,x_2\right]\nonumber\\
&\qquad-
\exp\left[\left(\tilde\beta-\frac{u}{2}\right)\,x_2\right]\Bigg\}\,\theta(-\tilde\beta).
\displaybreak[0]\\ \label{Hm01.90}
\tilde G_{2\,\mathrm{u}}(t;x_1,x_2)&=\frac{2}{\pi}\,\mathrm{Re}
\int_0^\infty\rd k\;\exp\left(\frac{u}{2}\,x_1
-k^2\,t\right)\,\sin\left(k\,x_1+\tan^{-1}\frac{k}{\tilde\beta}\right)\nonumber\\
&\quad\times\frac{1}{u\,(u+2\,\ri\,k)-2\,\gamma^2-2\,k^2}\nonumber\\
&\quad\;\times\Bigg\{\frac{\ri\,(\tilde\beta+u+\ri\,k)}{\tilde\beta+\sqrt{2\,\gamma^2+k^2}}\,
\exp\left[\left(\frac{u}{2}-\sqrt{2\,\gamma^2+k^2}\right)\,x_2-\ri\,\tan^{-1}\frac{k}{\tilde\beta}\right]
\nonumber\\
&\qquad-\ri\,\exp\left[-\left(\frac{u}{2}+\ri\,k\right)\,x_2
-\ri\,\tan^{-1}\frac{k}{\tilde\beta}\right]\Bigg\},\nonumber\\
&=\frac{1}{\pi}\,\int_{-\infty}^\infty\rd k\;\exp\left(\frac{u}{2}\,x_1
-k^2\,t\right)\,\sin\left(k\,x_1+\tan^{-1}\frac{k}{\tilde\beta}\right)\nonumber\\
&\quad\times\frac{1}{u\,(u+2\,\ri\,k)-2\,\gamma^2-2\,k^2}\nonumber\\
&\quad\;\times\Bigg\{\frac{\ri\,(\tilde\beta+u+\ri\,k)}{\tilde\beta+\sqrt{2\,\gamma^2+k^2}}\,
\exp\left[\left(\frac{u}{2}-\sqrt{2\,\gamma^2+k^2}\right)\,x_2-\ri\,\tan^{-1}\frac{k}{\tilde\beta}\right]
\nonumber\\
&\qquad-\ri\,\exp\left[-\left(\frac{u}{2}+\ri\,k\right)\,x_2
-\ri\,\tan^{-1}\frac{k}{\tilde\beta}\right]\Bigg\}.
\end{align}

For large values of $t$, the expressions for $G_2$ take a simpler for. It is seen
that if $\tilde\beta$ is negative, then $\tilde G_{2\,\mathrm{b}}$ dominates
$\tilde G_{2\,\mathrm{u}}$. So one can neglect $\tilde G_{2\,\mathrm{u}}$.
For nonnegative values of $\tilde\beta$, the right-hand side of (\ref{Hm01.90})
is dominated by the integrand for small values of $k$. If $\tilde\beta$
is zero, $\beta$ and $u$ are either both positive or both zero. One then
arrives at the following expressions for large $t$ behavior of $G_2$.
\begin{align}\label{Hm01.91}
G_2(t;x_1,x_2)&=\frac{-2\,\tilde\beta}
{u\,(u-2\,\tilde\beta)-2\,\gamma^2+2\,\tilde\beta^2}
\,\exp[\beta\,x_1+(\tilde\beta^2-\gamma^2)\,t]\nonumber\\
&\quad\times\Bigg\{
\frac{u}{\tilde\beta+\sqrt{2\,\gamma^2-\tilde\beta^2}}\,
\exp\left[\left(\frac{u}{2}-\sqrt{2\,\gamma^2-\tilde\beta^2}\right)\,x_2\right]\nonumber\\
&\qquad-
\exp\left[\left(\tilde\beta-\frac{u}{2}\right)\,x_2\right]\Bigg\},\nonumber\\
&\qquad\qquad\tilde\beta<0.\\ \nonumber\displaybreak[0]\\ \label{Hm01.92}
G_2(t;x_1,x_2)&=\frac{1}{\sqrt{\pi\,t}\,(u^2-2\,\gamma^2)}\,
\exp\left(-\frac{x_1^2}{4\,t}+\frac{u}{2}\,x_1-\gamma^2\,t\right)\nonumber\\
&\quad\times\Bigg\{\frac{u}{\sqrt{2\,\gamma^2}}\,
\exp\left[\left(\frac{u}{2}-\sqrt{2\,\gamma^2}\right)\,x_2\right]-
\exp\left(-\frac{u}{2}\,x_2\right)\Bigg\},\nonumber\\
&\qquad\qquad\tilde\beta=0,\quad u>0.\\ \nonumber\displaybreak[0]\\ \label{Hm01.93}
G_2(t;x_1,x_2)&=\frac{1}{\sqrt{4\,\pi\,t}\,(2\,\gamma^2)}\,\exp(-\gamma^2\,t)\nonumber\\
&\quad\times\Bigg\{\exp\left[-\frac{(x_1-x_2)^2}{4\,t}\right]+
\exp\left[-\frac{(x_1+x_2)^2}{4\,t}\right]\Bigg\},\nonumber\\
&\qquad\qquad\tilde\beta=0,\quad u=0.\\ \nonumber\displaybreak[0]\\
\label{Hm01.94}
G_2(t;x_1,x_2)&=\frac{1}{\sqrt{4\,\pi\,t}\,(2\,\gamma^2-u^2)}\,
\exp\left[\frac{u}{2}\,(x_1-x_2)-\gamma^2\,t\right]\nonumber\\
&\quad\times\Bigg\{\exp\left[-\frac{(x_1-x_2)^2}{4\,t}\right]-
\exp\left[-\frac{(x_1+x_2)^2}{4\,t}\right]\Bigg\},\nonumber\\
&\qquad\qquad\tilde\beta>0.
\end{align}
\section{Conclusion}
A one-dimensional model of diffusing particles on a line was studied,
where particles diffuse and drift, and there are also particle creation
(or annihilation). Particle creation and annihilation takes place at the
boundaries as well. It was shown that the system does tend to
a stationary state, provided the rate of particle creation in the bulk
(if there is particle creation, rather than annihilation in the bulk),
is smaller than a certain limit which is determined by the particle
annihilation rates at the boundaries, and the drift velocity:
the larger these, the larger would be the upper limit for
particle production rate which can be tolerated without making
the particle density blow up at large times. It was also shown
that the system can exhibit a dynamical phase transitions.
The control parameters for the dynamical phase transition are
the drift velocities and the absorbtion rates at the boundaries.
There are two phases: the fast phase in which the relaxation time
does not depend on the boundary rates, and the slow phase in which
the relaxation time does depend on the boundary terms. Finally,
the effect of a random particle creation in the bulk was also studied.
This was done by adding a noise $\eta$ to the equation for the density
$\rho$. The model thus obtained could be regarded as some
generalization of the Edwards-Wilkinson model, having in addition to
the usual diffusion and stochastic production, a general boundary condition
and a bulk annihilation (or creation) term, as well as a drift.
It was shown that in particular, the bulk annihilation rate, the drift velocity,
and the boundary annihilation rates determine the behavior of the stochastic
part of the density, with large time behaviors which undergo phase transitions
similar to the dynamical phase transitions observed in the no-noise
case. The one and two point functions corresponding to the density
were explicitly calculated in different regions of the parameter space.
\\[\baselineskip]
\textbf{Acknowledgement}:  This work was supported by
the research council of the Alzahra University.

\end{document}